\documentclass[prc,aps,floats,twocolumn,twoside,preprintnumbers,
superscriptaddress,floatfix,nofootinbib,showpacs]{revtex4}
\usepackage{graphicx,epsfig,pstricks,rotating}
\usepackage{bm}

\newcommand{\eqnb}{\begin{equation}}
\newcommand{\eqne}{\end{equation}}

\newcommand{\question}[1]{}

\def\slashchar#1{\setbox0=\hbox{$#1$}  % set a box for #1
   \dimen0=\wd0     % and get its size
   \setbox1=\hbox{/} \dimen1=\wd1  % get size of /
   \ifdim\dimen0>\dimen1   % #1 is bigger
      \rlap{\hbox to \dimen0{\hfil/\hfil}} % so center / in box
      #1     % and print #1
   \else     % / is bigger
      \rlap{\hbox to \dimen1{\hfil$#1$\hfil}} % so center #1
      /      % and print /
   \fi}      %

\begin{document}

\title{ $\eta'$ multiplicity and the Witten-Veneziano relation at 
finite temperature }
\author{S. Beni\' c}
\affiliation{Physics Department, Faculty of Science,
University of Zagreb,
Bijeni\v{c}ka c. 32, Zagreb 10000, Croatia}
\author{D. Horvati\'c}
\affiliation{Physics Department, Faculty of Science,
University of Zagreb,
Bijeni\v{c}ka c. 32, 
Zagreb 10000, Croatia}
\author{D. Kekez}
\affiliation{Rugjer Bo\v skovi\'c Institute, 
Bijeni\v{c}ka c. 54, Zagreb 10000, Croatia}
\author{D. Klabu\v{c}ar}
\email{klabucar@oberon.phy.hr}
\thanks{Corresponding author}
\thanks{Senior Associate of Abdus Salam ICTP, Trieste, Italy}
\affiliation{Physics Department, Faculty of Science,
University of Zagreb,
Bijeni\v{c}ka c. 32, 
Zagreb 10000, Croatia}
%\date{\today}

\begin{abstract}
\noindent
We discuss and propose the minimal generalization of 
the Witten-Veneziano relation to finite temperatures, 
prompted by STAR and PHENIX experimental results on the
multiplicity of $\eta'$ mesons. 
After explaining why these results show that 
the zero-temperature Witten-Veneziano relation cannot be 
straightforwardly extended to temperatures $T$ too close 
to the chiral restoration temperature $T_{\rm Ch}$ and beyond, 
we find the quantity which should replace, at $T>0$, 
the Yang-Mills topological susceptibility
appearing in the $T=0$ Witten-Veneziano relation,
in order to avoid the conflict 
with experiment at $T>0$. This is illustrated through 
concrete $T$-dependences of pseudoscalar meson masses
in a chirally well-behaved, Dyson-Schwinger approach, 
but our results and conclusions are of a more general nature
and, essentially, model-independent.
\end{abstract}

\maketitle

\section{Introduction}
\label{Introduction}

\noindent 
The ultrarelativistic heavy-ion collider facilities like RHIC at BNL
and LHC at CERN strive to produce a new form of hot QCD matter. 
The experiments show \cite{Adams:2005dq,Muller:2006ee} that it
has very intricate properties and presents a big challenge especially 
for theoretical understanding. While above the (pseudo)critical 
temperature $T_c\sim 170$ MeV 
this matter is often called the quark-gluon plasma (QGP), it cannot be 
a perturbatively interacting quark-gluon gas (as widely expected before 
RHIC results \cite{Adams:2005dq,Muller:2006ee}) until significantly 
higher temperatures $T \gg T_c$.
Instead, the interactions and correlations in the hot QCD matter are 
still strong (e.g., see Refs. \cite{Shuryak:2004tx,Stephanov:2007fk})
so that its more recent and more precise name is {\it strongly 
coupled QGP} (sQGP) \cite{Stephanov:2007fk}. One of its peculiarities
seems to be that strong correlations in the 
form of quark-antiquark ($q\bar q$) bound states and resonances still 
exist \cite{Shuryak:2004tx,Blaschke:2003ut} in the sQGP well above $T_c$.
In the old QGP paradigm, even deeply bound charmonium ($c\bar c$) states such 
as $J/\Psi$ and $\eta_c$ were expected to unbind at $T \approx T_c$,
but lattice QCD simulations of mesonic correlators now indicate  
they persist till around $2 T_c$ \cite{Datta:2002ck,Asakawa:2003re} 
or even above \cite{Shuryak:2005pp}.
Similar indications for light-quark mesonic bound states are also 
accumulating from lattice QCD \cite{Karsch:2002wv} and from other 
methods \cite{Shuryak:2004tx,Maris:2000ig,Mannarelli:2005pz}.
This agrees well with the findings on the lattice 
(e.g., see Ref. \cite{Fodor:2009ax} for a review) that for 
realistic explicit chiral symmetry breaking (ChSB), i.e., for the 
physical values of the current quark masses, the transition between
the hadron phase and the phase dominated by quarks and gluons, is
not an abrupt, singular phase transition but a smooth, analytic 
crossover around the {\it pseudo}critical temperature $T_c$. 
It is thus not too surprising that a clear
experimental signal of, e.g., deconfinement, 
is still hard to find and identify unambiguously.  

The most compelling signal for 
production of a new form of QCD matter, i.e., sQGP,
would be a restoration - in hot and/or dense matter - of the 
symmetries of the QCD Lagrangian which are broken in the vacuum.    
One of them is the [SU$_A$($N_f$) flavor] chiral symmetry, whose 
dynamical breaking results in light, (almost-)Goldstone 
pseudoscalar ($P$) mesons -- namely the octet 
$P = \pi^0,\pi^\pm,K^0, {\bar K}^0, K^\pm,\eta$,
as we consider all three light-quark flavors, $N_f=3$.
The second one 
is the U$_A$(1) symmetry. Its breaking by the non-Abelian axial 
Adler-Bell-Jackiw anomaly (`gluon anomaly' for short) makes 
the remaining pseudoscalar meson of the light-quark sector, the 
$\eta'$, much heavier, preventing its appearance as the ninth 
(almost-)Goldstone boson of dynamical chiral symmetry breaking 
(DChSB) in QCD.

The first experimental signature of a partial restoration of the U$_A$(1) 
symmetry seems to have been found in the $\sqrt{s_{N\!N}} = 200$ GeV central 
Au+Au reactions at RHIC. 
Namely, Cs\"org\H{o} {\it et al.} \cite{Csorgo:2009pa,Vertesi:2009wf}
analyzed combined data of PHENIX \cite{Adler:2004rq} and STAR \cite{Adams:2004yc} 
collaborations very robustly,
through six popular models for hadron multiplicities, and found that 
at 99.9\% confidence level, the $\eta'$ mass, which in the vacuum is
$M_{\eta'}=957.8$ MeV, is reduced by at least 200 MeV inside the fireball.
It is the sign of the disappearing 
contribution of the gluon axial anomaly to the $\eta'$ mass, 
which would drop to a value readily understood together with 
the (flavor-symmetry-broken) octet of 
$q\bar q'$ ($q,q'=u,d,s$) pseudoscalar mesons.
This is the issue of the ``return of the prodigal Goldstone boson" 
predicted \cite{Kapusta:1995ww}
as a signal of the U$_A$(1) symmetry restoration.

Another related but less obvious issue to which we want 
to draw attention, concerns the status, at $T>0$, of the famous 
Witten-Veneziano relation (WVR) \cite{Witten:1979vv,Veneziano:1979ec}
\begin{equation}
M_{\eta'}^2 + M_\eta^2 - 2 M_K^2 = \frac{6 \, \chi_{\rm Y\!M}}{f_\pi^2} \, 
\label{WittenVenez}
\end{equation}
between the $\eta'$, $\eta$ and $K$-meson masses $M_{\eta',\eta,K}$, 
pion decay constant $f_\pi$, and Yang-Mills (YM) 
topological susceptibility $\chi_{\rm Y\!M}$. 
WVR was obtained in the limit of large number of colors $N_c$
\cite{Witten:1979vv,Veneziano:1979ec}. It is well satisfied at $T=0$ for 
$\chi_{\rm Y\!M}$ obtained by lattice calculations (e.g.,
\cite{Lucini:2004yh,DelDebbio:2004ns,Alles:2004vi,Durr:2006ky}). 
Nevertheless,
the $T$-dependence of $\chi_{\rm Y\!M}$ is such \cite{Horvatic:2007qs} that 
the straightforward extension of Eq. (\ref{WittenVenez}) to $T>0$
\cite{Horvatic:2007qs},
i.e., replacement of all quantities{\footnote{Throughout this paper,
all quantities are for definiteness assumed at $T=0$ unless their
$T$-dependence is specifically indicated in formulas or in the text.}}
therein by their respective 
$T$-dependent versions $M_{\eta'}(T)$, $M_{\eta}(T)$, $M_K(T)$,
$f_\pi(T)$ {\it and} $\chi_{\rm Y\!M}(T)$, leads to a conflict with 
experiment \cite{Csorgo:2009pa,Vertesi:2009wf}.
Since this extension of Eq. (\ref{WittenVenez}) to $T>0$
was studied in Ref. \cite{Horvatic:2007qs} before the pertinent
experimental analysis \cite{Csorgo:2009pa,Vertesi:2009wf}, one of 
the purposes of this paper is to revisit the implications of the 
results of Ref. \cite{Horvatic:2007qs} for WVR at $T>0$, and 
demonstrate explicitly that they are practically model-independent.
The other, more important purpose is to propose a mechanism
which can enable WVR to agree with experiment at $T>0$.

\section{The relations connecting two theories, QCD and YM }
\label{relationsConnect}

\noindent Both issues pointed out before Eq. (\ref{WittenVenez}) and 
around it, are best understood in a model-independent way if 
one starts from the chiral limit of vanishing current quark masses 
($m_q = 0$) for all three light %quark 
flavors, $q = u, d, s$. 
Then not only pions and kaons are massless, but is also  
$\eta$, which is then (since the situation is also 
SU(3)-flavor-symmetric) a purely SU(3)-octet state, $\eta=\eta_8$. 
In contrast, $\eta'$ is then purely singlet, $\eta'=\eta_0$;
since the divergence of the singlet axial quark current 
${\bar q}\gamma^\mu \gamma_5 {\frac{1}{2}} \lambda^0 q$ is nonvanishing 
even for $m_q = 0$ due to the gluon anomaly, the $\eta'$ mass squared
receives the anomalous contribution $\Delta M_{\eta'}^2$ 
($=\lambda^4/f_{\eta'}^2$ in the notation of Ref. \cite{Kapusta:1995ww})
which is nonvanishing even in the chiral limit:
\begin{equation}
\frac{\lambda^4}{f_{\eta'}^2}=\Delta M_{\eta_0}^2=\Delta M_{\eta'}^2 
= \frac{6 \, \chi_{\rm Y\!M}}{f_{\pi}^2} + O(\frac{1}{N_c}).
\label{WV_ChLim}
\end{equation}
However, $\lambda^4 $ and $f_{\eta'}$ are known accurately{\footnote{Also note
that a unique ``$\eta'$ decay constant" $f_{\eta'}$ is, strictly speaking, 
not a well-defined quantity, as {\it two} $\eta'$ decay constants are actually 
needed: the singlet one, $f^0_{\eta'}$, and the octet one, $f^8_{\eta'}$;
{\it e.g.}, see an extensive review \cite{Feldmann:1999uf} or  
the short Appendix of Ref. \cite{Kekez:2000aw}.}}
only in the 
large $N_c$ limit. There, in the leading order in $1/N_c$, $\lambda^4$ 
is given by the YM (i.e., ``pure glue") topological susceptibility 
$\chi_{\rm Y\!M}$ times $2N_f=6$ \cite{Witten:1979vv,Veneziano:1979ec}, 
and the ``$\eta'$ decay constant" $f_{\eta'}$ is the same as 
$f_{\pi}$ \cite{Leutwyler:1992yt}.
Thus, keeping only the leading order in $1/N_c$, the last equality
is WVR in the chiral limit.

The consequences of Eq. (\ref{WV_ChLim}) remain qualitatively the 
same realistically away from the chiral limit. This will soon 
become clear on the basis of, e.g., Eq. (\ref{M2_prop_m}) below.  
Namely, due to DChSB in QCD, 
for relatively light current quark masses $m_q$ $(q=u,d,s)$, 
the $q\bar q'$ bound-state pseudoscalar meson masses (including 
the {\it nonanomalous} parts of the $\eta'$ and $\eta$ masses)  
behave as 
\begin{equation}
M_{q\bar q'}^2 = {\rm const}\, ({m}_q +{m}_{q'})
 \, ,  \quad (q, q' = u,d,s).
\label{M2_prop_m}
\end{equation}
The pseudoscalar mesons (including $\eta'$) thus obtain 
relatively light nonanomalous contributions $M_{q\bar q'}$
to their masses $M_P$, allowing them to reach the empirical values.
That is, instead of the eight strictly massless Goldstone bosons,
$\pi^0,\pi^\pm,K^0, {\bar K}^0, K^\pm$ and $\eta$ are relatively 
light almost-Goldstones. Among them, in the limit of isospin symmetry
($m_u\! =\! m_d$), only $\eta$ now receives also the gluon-anomaly 
contribution since the explicit SU(3) flavor breaking 
between the nonstrange ($NS$) $u,d$-quarks and $s$-quarks 
causes the mixing between the isoscalars $\eta$ and $\eta'$. 
For $m_q\neq 0$, Eq. (\ref{WV_ChLim}) is replaced by
the usual WVR (\ref{WittenVenez}) containing also the nonanomalous 
contributions to meson masses. Nevertheless, these contributions
largely cancel due to the approximate SU(3) flavor symmetry and
to DChSB [i.e., Eq. (\ref{M2_prop_m})].

This can be seen assuming the usual SU(3) $q\bar q$ content of the
pseudoscalar meson nonet with well-defined isospin{\footnote{Vanishing of 
the anomaly at sufficiently high $T$ opens the possibility of 
studying the interesting scenario of maximal isospin violation at high $T$
\cite{Kharzeev:1998kz,Kharzeev:2000na}, but as the effects of the
small difference between $m_u$ and $m_d$ are not important
for the present considerations, we stick to the isospin limit
throughout the present paper.}} quantum numbers, 
in particular the isoscalar $(I=0)$ octet and singlet etas,
$\eta_8 = (u\bar{u} + d\bar{d} - 2s\bar{s})/\sqrt{6}$,
$\eta_0 = (u\bar{u} + d\bar{d} + s\bar{s})/\sqrt{3}$,
whose mixing yields the physical particles $\eta$ and $\eta'$.
Since the {\it nonanomalous} parts of the $\eta_0$ and $\eta_8$
masses squared, $M_{\eta_0}^2$ and $M_{\eta_8}^2$, are respectively
$M_{00}^2 \approx \frac{2}{3} M_K^2 + \frac{1}{3} M_\pi^2$ and 
$M_{88}^2 \approx \frac{4}{3} M_K^2 - \frac{1}{3} M_\pi^2$ 
(see, e.g., Ref. \cite{Kekez:2005ie}), and since
$M_{\eta_8}^2 + M_{\eta_0}^2 = M_{\eta}^2 + M_{\eta'}^2$,
the nonanomalous parts of the $\eta$ and $\eta'$ masses
are canceled by $2 M_K^2$ in WVR (\ref{WittenVenez}).
Another way of seeing this is expressing the nonanomalous 
parts of $M_{\eta}^2 + M_{\eta'}^2 = M_{\eta_8}^2 + M_{\eta_0}^2$
by Eq. (\ref{M2_prop_m}).        Thus again 
$M_{\eta}^2 + M_{\eta'}^2 - 2 M_K^2 \approx \Delta M_{\eta_0}^2$,
showing again that already WVR's chiral-limit-nonvanishing part 
(\ref{WV_ChLim}) reveals the essence of the influence of the gluon anomaly 
on the masses in the $\eta'$-$\eta$ complex. 
This is important also for the presently pertinent finite-$T$ context 
because thanks to this, below it will be shown model-independently that 
WVR (\ref{WittenVenez}) 
containing the YM topological susceptibility $\chi_{\rm Y\!M}$ implies 
$T$-dependence of $\eta'$ mass in conflict with the recent experimental 
results \cite{Csorgo:2009pa,Vertesi:2009wf}.

Namely, the gluon-anomaly contribution (\ref{WV_ChLim}) 
is established at $T=0$
but it is not expected to persist at high temperatures. Ultimately,
$\eta'$ should also become a massless Goldstone boson at sufficiently
high $T$, where $\chi_{\rm Y\!M}(T) \to 0$.
However, according to WVR, $\Delta M_{\eta'}(T)$ falls only for $T$ where
$f_\pi(T)^2$ does not fall faster than $6\chi_{\rm Y\!M}(T)$,
as stressed in Ref. \cite{Horvatic:2007qs}.  

The WVR's chiral-limit version (\ref{WV_ChLim}) manifestly
points out the ratio $\chi_{\rm Y\!M}(T)/f_\pi(T)^2$ as crucial for the 
anomalous $\eta'$ mass, but the above discussion shows that this remains
essentially the same away from the chiral limit.

In the present context, it is important for practical calculations
to go realistically away from the chiral limit, in which the chiral
restoration is a sharp phase transition at its critical temperature
$T_{\rm Ch}$ where the chiral-limit pion decay constant vanishes
very steeply, i.e., as steeply as the chiral quark condensate.
In contrast, for realistic explicit ChSB, i.e., $m_u$ and $m_d$ of
several MeV, this transition is a {\it smooth crossover} (e.g., see Ref.
\cite{Fodor:2009ax}).  For the pion decay constant, this implies that
$f_\pi(T)$ still falls relatively steeply around {\it pseudo}critical
temperature $T_{\rm Ch}$, but less so than in the chiral case, and even
remains finite, enabling the usage of WVR (\ref{WittenVenez}) for the
temperatures across the chiral and U$_A$(1) symmetry restorations.

WVR is very remarkable because it connects two different theories: 
QCD with quarks and its pure-gauge, YM counterpart. The latter, however, 
has much higher characteristic temperatures than QCD with quarks: the 
``melting temperature" $T_{\rm Y\!M}$ where $\chi_{\rm Y\!M}(T)$ 
starts to decrease appreciably was found on lattice to be, for example,
$T_{\rm Y\!M} \approx 260$ MeV \cite{Alles:1996nm,Boyd:1996bx} or 
even higher, $T_{\rm Y\!M} \approx 300$ MeV \cite{Gattringer:2002mr}.
In contrast, the pseudocritical temperatures for the chiral and
deconfinement transitions in the full QCD are lower than $T_{\rm Y\!M}$ 
by some 100 MeV or more (e.g., see Ref. \cite{Fodor:2009ax}) 
due to the presence of the quark degrees of freedom. 

\begin{figure}[b]
\centerline{\includegraphics[width=80mm,angle=0]{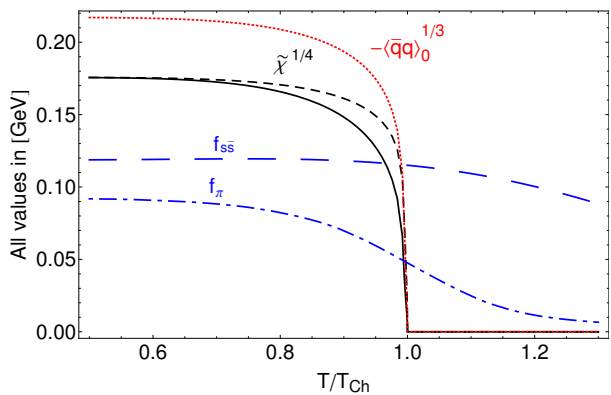}}
\caption{The relative-temperature dependences, on $T/T_{\rm Ch}$, of
${\widetilde \chi}^{1/4}$, $\langle{\bar q}q\rangle_0^{1/3}$,
$f_\pi$ and $f_{s\bar s}$, i.e.,  
the $T/T_{\rm Ch}$-dependences of the quantities entering in the
anomalous contributions to various masses in the $\eta'$-$\eta$
complex -- see Eq. (\ref{MetaPrime}) and formulas below it.
The solid curve depicts ${\widetilde \chi}^{1/4}$ for $\delta=0$
in Eq. (\ref{chitildeChiCond}), and the short-dashed curve is
${\widetilde \chi}^{1/4}$ for $\delta=1$.  At $T=0$, the both
${\widetilde \chi}\,$'s are equal to $\chi_{\rm Y\!M} = (0.1757 
\, \rm GeV)^4$, the weighted average \cite{Horvatic:2007qs}
of various lattice results for $\chi_{\rm Y\!M}$.
% Color online. 
The dotted (red) curve depicts $ - \langle{\bar q}q\rangle_0^{1/3}$,
the dash-dotted (blue) curve is $f_\pi$, and the long-dashed
(blue) curve is $f_{s\bar s}$. (Colors online.)
}
\label{chi1757_qq_f}
\end{figure}

This difference in characteristic temperatures, in conjunction with 
$\chi_{\rm Y\!M}(T)$ in WVRs (\ref{WittenVenez}) and (\ref{WV_ChLim}) 
would imply that the (partial) restoration of the U$_A$(1) symmetry 
(understood as the disappearance of the anomalous $\eta_0/\eta'$ mass)
should happen well after the restoration of the chiral symmetry. 
But, this contradicts the RHIC experimental observations of the 
reduced $\eta'$ mass \cite{Csorgo:2009pa,Vertesi:2009wf} {\it if} 
WVRs (\ref{WittenVenez}), (\ref{WV_ChLim}) hold unchanged also close to 
the QCD chiral restoration temperature $T_{\rm Ch}$, around which 
$f_\pi(T)$ decreases still relatively steeply{\footnote{Relative to 
decay constants of mesons containing a strange quark; e.g., compare 
$f_{s\bar s}(T)$ of the unphysical ${s\bar s}$ pseudoscalar 
with $f_\pi(T)$ in Fig. \ref{chi1757_qq_f}.}} 
\cite{Horvatic:2007qs} for realistic explicit ChSB,
thus leading to the increase of $6 \chi_{\rm Y\!M}(T)/f_\pi(T)^2$ and 
consequently also of $M_{\eta'}$.

There is still more to the relatively high resistance of 
$\chi_{\rm Y\!M}(T)$ to temperature: not only does it start falling
at rather high $T_{\rm Y\!M}$, but $\chi_{\rm Y\!M}(T)$ 
found on the lattice is falling with $T$ {\it relatively} slowly. 
In some of the applications in the past (e.g., see Refs. 
\cite{Fukushima:2001hr,Schaffner-Bielich:1999uj}), it was 
customary to simply rescale a temperature characterizing the 
pure-gauge, YM sector to a value characterizing QCD 
with quarks. (For example, Refs. 
\cite{Fukushima:2001hr,Schaffner-Bielich:1999uj} rescaled 
$T_{\rm Y\!M} = 260$ MeV found by Ref. \cite{Alles:1996nm} to 150 MeV). 
However, even if we rescale the critical temperature for melting
of the topological susceptibility $\chi_{\rm Y\!M}(T)$ from 
%the original value of 
$T_{\rm Y\!M}$ down to $T_{\rm Ch}$, the value of $6 \chi_{\rm Y\!M}(T)/f_\pi(T)^2$ 
still increases a lot \cite{Horvatic:2007qs} for 
the pertinent temperature interval starting already below $T_{\rm Ch}$. 
This happens because $\chi_{\rm Y\!M}(T)$ falls with $T$ more slowly 
than $f_\pi(T)^2$. (It was found \cite{Horvatic:2007qs}
that the rescaling of $T_{\rm Y\!M}$ would have to be 
totally unrealistic, to less than 70\% of $T_{\rm Ch}$, in order to 
achieve sufficiently fast drop of the anomalous 
contribution that would allow the observed enhancement in the 
$\eta'$ multiplicity.)

These WVR-induced enhancements of the $\eta'$ mass for $T \sim T_{\rm Ch}$ 
were first noticed in Ref. \cite{Horvatic:2007qs}. This reference used a 
concrete dynamical model (with an effective, rank-2 separable interaction,
convenient for computations at $T\geq 0$) \cite{Blaschke:2000gd} of low energy, 
nonperturbative QCD to obtain mesons as $q\bar q'$ bound states in Dyson-Schwinger 
(DS) approach \cite{e.g.reviews,Roberts:2000aa,Alkofer:2000wg}, which is a 
bound-state approach with the correct chiral behavior (\ref{M2_prop_m}) 
of QCD.  Nevertheless, this concrete dynamical DS model was used in Ref. 
\cite{Horvatic:2007qs} to get concrete values for only the nonanomalous 
parts of the meson masses, but was essentially {\it not} used to 
get model predictions for the mass contributions from the gluon anomaly, 
in particular $\chi_{\rm Y\!M}(T)$. On the contrary, the anomalous mass 
contribution was included, in the spirit of $1/N_c$ expansion, through 
WVR (\ref{WittenVenez}). Thus, the $T$-evolution of the $\eta'$-$\eta$ 
complex in Ref. \cite{Horvatic:2007qs} was not dominated by dynamical model 
details, but by WVR, i.e., the ratio $6 \chi_{\rm Y\!M}(T)/f_\pi(T)^2$. 
 Admittedly, $f_\pi(T)$ was also calculated within this model,
causing some {\it quantitative} model dependence of the anomalous mass
in WVR, but this cannot change the qualitative observations of Ref. 
\cite{Horvatic:2007qs} on the $\eta'$ mass enhancement. Namely, our model 
$f_\pi(T)$, depicted as the dash-dotted curve in Fig. \ref{chi1757_qq_f}, 
obviously has the right crossover features \cite{Fodor:2009ax}. It also
agrees qualitatively with $f_\pi(T)$'s calculated in other realistic 
dynamical models \cite{Maris:2000ig,Roberts:2000aa}. 
 Various modifications were tried in Ref. \cite{Horvatic:2007qs} 
but could not reduce much the $\eta'$ mass enhancement caused by this 
ratio, let alone bring about the significant $\eta'$ mass reduction 
found in the RHIC experiments \cite{Csorgo:2009pa,Vertesi:2009wf}.

One must therefore conclude that either WVR breaks down 
as soon as $T$ approaches $T_{\rm Ch}$, or that the $T$-dependence
of its anomalous contribution is different from the pure-gauge
$\chi_{\rm Y\!M}(T)$.
We will show that the latter alternative is possible, since WVR 
can be reconciled with experiment thanks to the existence of 
another relation which, similarly to WVR, connects the YM theory 
with full QCD.  Namely, using large-$N_c$ arguments, 
Leutwyler and Smilga derived \cite{Leutwyler:1992yt}, at $T=0$, 
%the relation 
\begin{equation}
\chi_{\rm Y\!M} = 
\frac{ \chi }{1 + \chi\, \frac{N_f}{m\,\langle{\bar q}q\rangle_0} }
\, \left( \equiv {\widetilde \chi} \right)
 \, ,
\label{chitilde}
\end{equation}
the relation (in our notation) between the YM topological susceptibility 
$\chi_{\rm Y\!M}$, and the full-QCD topological susceptibility $\chi$,
the {\it chiral-limit} quark condensate 
$\langle{\bar q}q\rangle_0$, and $m$, the harmonic average of $N_f$ 
current quark masses $m_q$. That is, $m$ is $N_f$ times the reduced mass.  
In the present case of $N_f=3$, $q=u,d,s$, so that
\begin{equation}
\frac{N_f}{m} \, = 
\,  {\displaystyle\sum\limits_{q=u,d,s}}\,\, \frac{1}{m_q}  \, .
\end{equation}

%Leutwyler-Smilga relation 
Equation (\ref{chitilde}) is a remarkable relation between the two pertinent theories.
For example, in the limit of all very heavy quarks ($m_q\to \infty,\, q=u,d,s $), 
it correctly leads to the result that $\chi_{\rm Y\!M}$ is equal to the value 
of the topological susceptibility in {\it quenched} QCD,
$\chi_{\rm Y\!M} = \chi(m_q=\infty)$. This holds because $\chi$ 
is by definition the vacuum expectation value of a gluonic operator, so that
the absence of quark loops would leave only the pure-gauge, YM contribution. 
However, the Leutwyler-Smilga relation (\ref{chitilde}) also holds 
in the opposite (and presently pertinent) limit of light quarks. 
This limit still presents a problem for getting the full-QCD
topological susceptibility $\chi$ on the lattice \cite{DeGrand:2007tx},
but we can use the light-quark-sector
result \cite{Di Vecchia:1980ve,Leutwyler:1992yt} 
%that for small $m_q$,
\begin{equation}
\chi = - \frac{m\,\langle{\bar q}q\rangle_0}{N_f} + {\cal C}_m
 \, ,
\label{chi_small_m}
\end{equation}
where ${\cal C}_m$ stands for corrections of higher orders in
small $m_q$, and thus of small magnitude.  The leading term
is positive (as $\langle{\bar q}q\rangle_0 <0$), but
${\cal C}_m$ is negative, since Eq. (\ref{chitilde}) shows that
$\chi \leq {\rm min}(- m\,\langle{\bar q}q\rangle_0/N_f, \chi_{\rm Y\!M})$.

Although small, ${\cal C}_m$ should not be neglected, since 
${\cal C}_m = 0$ would imply, through Eq. (\ref{chitilde}), 
that $\chi_{\rm Y\!M} = \infty$. 
Instead, its value (at $T=0$) is fixed by Eq. (\ref{chitilde}): 
\begin{equation}
{\cal C}_m = {\cal C}_m(0) = \frac{m\,\langle{\bar q}q\rangle_0}{N_f}\, \left( 
1 - \chi_{\rm Y\!M}\, \frac{N_f}{m\,\langle{\bar q}q\rangle_0}\right)^{-1} \, .
\label{Cat0}
\end{equation}

All this starting from Eq. (\ref{chitilde}) has so far been at $T=0$. 
If the left- and right-hand side of Eq. (\ref{chitilde})
are extended to $T>0$, it is obvious that the equality cannot hold 
at arbitrary temperature $T>0$. The relation (\ref{chitilde}) must 
break down somewhere close to the (pseudo)critical temperatures
of full QCD ($\sim T_{\rm Ch}$) since the pure-gauge quantity 
$\chi_{\rm Y\!M}$ is much more temperature-resistant than the right-hand side,
abbreviated as ${\widetilde \chi}$. The quantity ${\widetilde \chi}$, 
which may be called the effective susceptibility, consists of the 
full-QCD quantities $\chi$ and $\langle{\bar q}q\rangle_0$,
the quantities of full QCD with quarks, characterized by 
$T_{\rm Ch}$, just as $f_\pi(T)$.  As $T\to T_{\rm Ch}$, the 
chiral quark condensate $\langle{\bar q}q\rangle_0(T)$ drops faster 
than the other DChSB parameter in the present problem, namely $f_\pi(T)$ 
for realistically small explicit ChSB. (See Fig. \ref{chi1757_qq_f} 
for the results of the dynamical model adopted here from Ref. 
\cite{Horvatic:2007qs}, and, e.g., Refs. \cite{Maris:2000ig,Roberts:2000aa} 
for analogous results of different DS models).
Thus, the troublesome mismatch in $T$-dependences of $f_\pi(T)$
and the pure-gauge quantity $\chi_{\rm Y\!M}(T)$, which causes 
the conflict of the temperature-extended WVR with experiment 
at $T \agt T_{\rm Ch}$, is expected to disappear if 
$\chi_{\rm Y\!M}(T)$ is replaced by  ${\widetilde \chi(T)}$,
the temperature-extended effective susceptibility.
The successful zero-temperature WVR (\ref{WittenVenez}) is, however,  
retained, since $\chi_{\rm Y\!M} = {\widetilde \chi}$ at $T=0$.

Extending Eq. (\ref{chi_small_m}) to $T>0$ is something of a 
guesswork as there is no guidance from the lattice for $\chi(T)$ 
[unlike $\chi_{\rm Y\!M}(T)$]. Admittedly, the leading term is 
straightforward as it is plausible that its $T$-dependence will 
simply be that of $\langle{\bar q}q\rangle_0(T)$. Nevertheless, 
for the correction term ${\cal C}_m$ such a plausible assumption 
about the form of $T$-dependence cannot be made and Eq. (\ref{Cat0}), 
which relates YM and QCD quantities, only gives its value at $T=0$.
We will therefore explore the $T$-dependence of the anomalous masses 
using the following Ansatz for the $T\geq 0$ generalization of 
Eq. (\ref{chi_small_m}):
\begin{equation}
\chi(T) = - \frac{m\,\langle{\bar q}q\rangle_0(T)}{N_f} + {\cal C}_m(0) \, 
\left[\frac{\langle{\bar q}q\rangle_0(T)}{\langle{\bar q}q\rangle_0(T=0)}\right]^{\delta}
 \, ,
\label{chi_small_mT}
\end{equation}
where the correction-term $T$-dependence is parametrized through the power 
%$0\leq\delta\leq 2$ 
$\delta$ of the presently fastest-vanishing (as $T \to T_{\rm Ch}$) 
chiral order parameter $\langle{\bar q}q\rangle_0(T)$.

\begin{figure}[b]
\centerline{\includegraphics[width=80mm,angle=0]{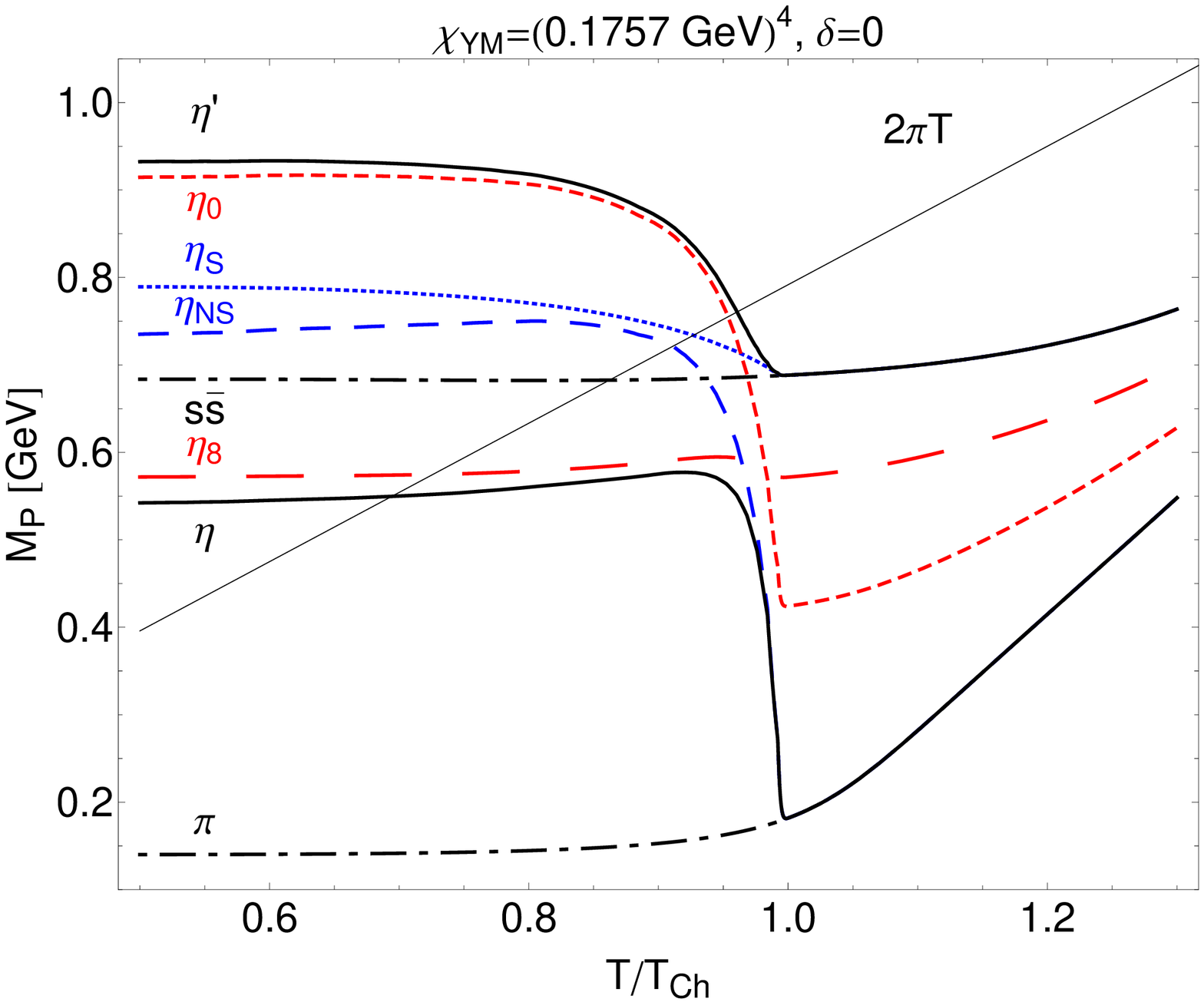}}
\centerline{\includegraphics[width=80mm,angle=0]{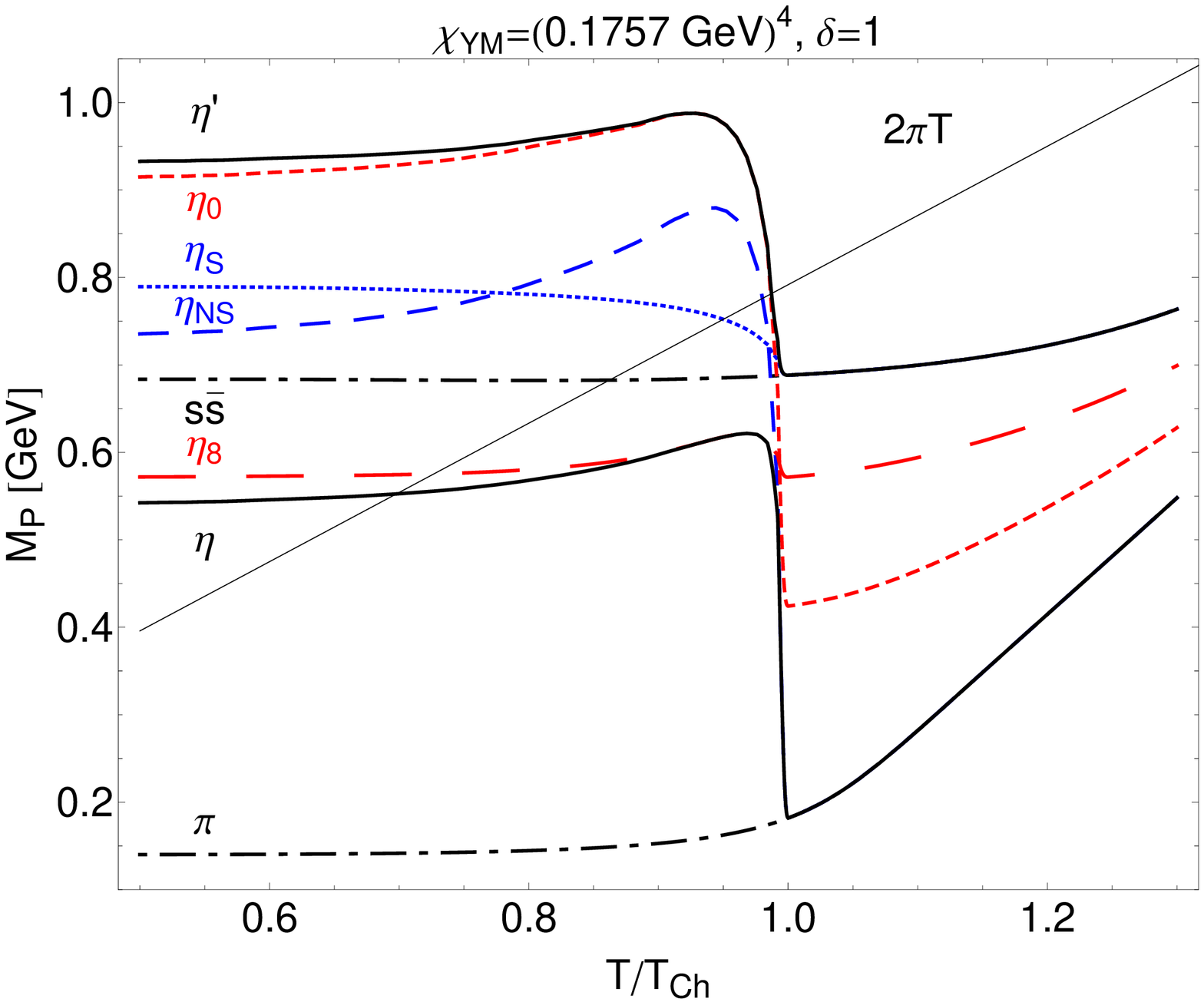}}
\caption{The relative-temperature dependence, on $T/T_{\rm Ch}$,
of the pseudoscalar meson masses for two ${\widetilde \chi}(T)$,
namely Eq. (\ref{chitildeChiCond}) with $\delta=0$ (upper panel)
and with $\delta=1$ (lower panel). 
The meaning of all symbols is the same on the both panels: the 
masses of $\eta'$ and $\eta$ are, respectively, the upper and 
lower solid curve, those of the pion and nonanomalous $s\bar s$ 
pseudoscalar are, respectively, the lower and upper dash-dotted 
curve, $M_{\eta_0}$ and $M_{\eta_8}$ are, respectively, the
short-dashed (red) and long-dashed (red) curve, $M_{\eta_{NS}}$ is 
the medium-dashed (blue), and $M_{\eta_{S}}$ is the dotted (blue) 
curve. (Colors online.)
The straight line $2\pi T$ is twice the lowest Matsubara frequency.}
\label{M_Pchi1757}
\end{figure}

The $T\geq 0$ extension (\ref{chi_small_mT}) of the 
light-quark $\chi$, Eq.~(\ref{chi_small_m}), leads to 
the $T\geq 0$ extension of ${\widetilde \chi}$:
\begin{equation}
\nonumber
{\widetilde \chi}(T) = \frac{m\,\langle{\bar q}q\rangle_0(T)}{N_f}
\left(1 - \frac{m\,\langle{\bar q}q\rangle_0(T)}{N_f \, {\cal C}_m(0)}
\left[\frac{\langle{\bar q}q\rangle_0(T=0)}{\langle{\bar q}q\rangle_0(T)}\right]^{\delta}
\right).
\label{chitildeChiCond}
\end{equation}
We now use ${\widetilde \chi}(T)$ in WVR instead of 
$\chi_{\rm Y\!M}(T)$ used by Ref. \cite{Horvatic:2007qs}.
This gives us the temperature dependences of the masses 
in the $\eta$-$\eta'$ complex, such as those in Fig. \ref{M_Pchi1757}
illustrating the cases $\delta=0$ and $\delta=1$.

It is clear that ${\widetilde \chi}(T)$ (\ref{chitildeChiCond}) 
blows up as $T \to T_{\rm Ch}$ if the correction term there 
vanishes faster than $\langle{\bar q}q\rangle_0(T)$ {\it squared}.
Thus, varying $\delta$
between 0 and 2 covers the cases from the $T$-independent correction term,
to (already experimentally excluded) enhanced anomalous masses for $\delta$ 
noticeably above 1, to even sharper mass blow-ups for $\delta\to 2$ 
when $T \to T_{\rm Ch}$.
On the other hand, it does not seem natural that the correction term vanishes 
faster than the fastest-vanishing order parameter $\langle{\bar q}q\rangle_0(T)$. Indeed, already for the same rate of vanishing of the both terms
($\delta=1$),
one can notice in Fig. \ref{M_Pchi1757} the start of the precursors of the 
blow-up of various masses in the $\eta'$-$\eta$ complex as $T \to T_{\rm Ch}$
although these small mass bumps are still experimentally acceptable. 
Thus, in Fig. \ref{M_Pchi1757} we depict the $\delta=1$ case, and $\delta=0$
($T$-independent correction term) as the other acceptable extreme. Since
they turn out to be not only qualitatively, but also quantitatively so
similar that the present era experiments cannot discriminate between them,
there is no need to present any `in-between results', for $0 <\delta < 1$.
 
Next we turn to completing the explanation how the above-mentioned results 
in Fig. \ref{M_Pchi1757} were obtained. 

%To clarify completely how the results in Fig. \ref{M_Pchi1757} 
%were obtained, we now give some additional explanations.

Using ${\widetilde \chi}(T)$ in WVR instead of $\chi_{\rm Y\!M}(T)$ 
used by Ref. \cite{Horvatic:2007qs}, does not change anything at
$T=0$, where ${\widetilde \chi}(T) = \chi_{\rm Y\!M}(0)$,
which remains an excellent approximation even well beyond $T=0$.
Nevertheless, this changes drastically as $T$ approaches $T_{\rm Ch}$.
For $T \sim T_{\rm Ch}$, the behavior of ${\widetilde \chi}(T)$
is dominated by the $T$-dependence of the chiral condensate,
tying the restoration of the U$_A$(1) symmetry to the 
chiral symmetry restoration.

As for the nonanomalous contributions to the meson masses,
we use the same DS model 
(and parameter values) as in Ref. \cite{Horvatic:2007qs}, 
since it includes both DChSB and correct QCD chiral behavior 
as well as realistic explicit ChSB. 
That is, all nonanomalous results 
($M_\pi, f_\pi, M_K, f_K$, as well as $M_{s\bar s}$ and 
$f_{s\bar s}$, the mass and decay constant of the unphysical
$s\bar s$ pseudoscalar meson) in the present paper are, for all
$T$, taken over from Ref. \cite{Horvatic:2007qs}.  We used 
this same model also for computing the chiral quark condensate 
$\langle{\bar q}q\rangle_0$, including its $T$-dependence 
displayed in Fig. \ref{chi1757_qq_f}.

This defines completely how the results displayed in Fig. \ref{M_Pchi1757}
were generated. For details, see Ref. \cite{Horvatic:2007qs} (and Ref. 
\cite{Kekez:2005ie} for $M_{\eta_0}$ and $M_{\eta_8}$). Here we list only 
the formulas which, in conjunction with Fig. \ref{chi1757_qq_f}, enable 
the reader to understand easily the $T$-dependences of the masses in 
Fig. \ref{M_Pchi1757}:
The theoretical $\eta'$ and $\eta$ mass eigenvalues are
\begin{eqnarray}
M_{\eta'}^2(T) &=& \frac{1}{2} \left[ M_{\eta_{NS}}^2(T) + M_{\eta_{S}}^2(T)
                           + \Delta_{\eta \eta'}(T) \right] , \quad
\label{MetaPrime}
\\
\label{Meta}
M_{\eta}^2(T) &=& \frac{1}{2} \left[ M_{\eta_{NS}}^2(T) + M_{\eta_{S}}^2(T)
                          - \Delta_{\eta \eta'}(T) \right] , \quad
\end{eqnarray}
where $\Delta_{\eta \eta'} \equiv \sqrt{[M_{\eta_{NS}}^2  -
M_{\eta_{S}}^2]^2 + 8 \beta^2 X^2}$ ,     
$$  \beta = \frac{1}{2+X^2}\, \frac{6\, {\widetilde \chi}}{f_\pi^2} \, , 
\qquad         X \equiv \frac{f_\pi}{f_{s\bar s}}  \, ,  $$                       
$$M_{\eta_{NS}}^2 = M_\pi^2 + 2\beta \, , \qquad
M_{\eta_S}^2 = M_{s\bar s}^2 + \beta X^2  \, ,   $$
$$M_{\eta_0}^2 = M_{00}^2 + \frac{1}{3}(2+X)^2 \, \beta \, , \, 
M_{\eta_8}^2 = M_{88}^2 + \frac{2}{3}(1-X)^2 \, \beta  , 
$$
$$ M_{88}^2 = \frac{2}{3}\, M_{s\bar{s}}^2 + \frac{1}{3}\, M_{\pi}^2 \,\,\, , 
\,\,\,
M_{00}^2 = \frac{1}{3}\, M_{s\bar{s}}^2 + \frac{2}{3} \, M_{\pi}^2 \, . $$
In all expressions after Eq. (\ref{Meta}), the $T$-dependence is understood. 

In both cases considered for the                  
topological susceptibility (\ref{chi_small_mT})
[$\delta=0$, i.e., the constant correction term, and $\delta=1$, i.e.,
the strong $T$-dependence $\propto \langle{\bar q}q\rangle_0(T)$
of both the leading and correction terms in $\chi(T)$],
the results are consistent with the experimental findings
on the decrease of the $\eta'$ mass of
Cs\"org\H{o} {\it et al.} \cite{Csorgo:2009pa,Vertesi:2009wf}.

\section{Summary, discussion and conclusions}
\label{conclusions}

\noindent 
In the light of the recent experimental results on the $\eta'$ 
multiplicity in heavy-ion collisions \cite{Csorgo:2009pa,Vertesi:2009wf},
we revisited the earlier theoretical work \cite{Horvatic:2007qs} 
concerning the thermal behavior of the $\eta'$-$\eta$ complex
following from WVR straightforwardly extended to $T>0$.
We have confirmed the results of Ref. \cite{Horvatic:2007qs}
on WVR where the ratio $\chi_{\rm Y\!M}(T)/f_\pi(T)^2$
dominates the $T$-dependence, and clarified that these results
are practically model-independent.
It is important to note the difference between our approach 
and those that attempt to give model predictions for topological 
susceptibility, such as Refs. \cite{Contrera:2009hk,Costa:2008dp}.
By contrast, in 
Refs. \cite{Horvatic:2007qs} and here, as well as earlier works 
\cite{Klabucar:1997zi,Kekez:2000aw,Kekez:2001ph,Kekez:2005ie} at
$T=0$, a DS dynamical model is used (as far as masses are concerned) 
to obtain only the nonanomalous part 
of the light pseudoscalar meson masses (where the model dependence 
is however dominated by their almost-Goldstone character), while
the anomalous part of the masses in the $\eta'$-$\eta$ complex is, 
through WVR, dictated by $6\chi_{\rm Y\!M}/f_\pi^2$. In this ratio, 
$f_\pi(T)$ is admittedly model-dependent in quantitative sense, but 
other realistic models yield qualitatively similar crossover 
behaviors~\cite{footnote5} of $f_\pi(T)$ for $m_q \neq 0$, as exemplified 
by our Fig. \ref{chi1757_qq_f}, and Fig. 2 in Ref. \cite{Horvatic:2007qs}, 
and by Fig. 6 in Ref. \cite{Maris:2000ig}. Such $f_\pi(T)$ behaviors are 
also in agreement with the $T$-dependence expected of the DChSB order
parameter on general grounds: a pronounced fall-off around $T_{\rm Ch}$
-- but exhibiting, in agreement with lattice \cite{Fodor:2009ax},
a smooth crossover pattern for nonvanishing explicit ChSB,
a crossover which gets slower with growing $m_q$ [e.g., compare
$f_\pi(T)$ with $f_{s\bar s}(T)$ in Fig. \ref{chi1757_qq_f}].
In contrast to the QCD topological susceptibility $\chi$, the 
YM topological susceptibility and even its $T$-dependence
$\chi_{\rm Y\!M}(T)$, including its ``melting" temperature 
$T_{\rm YM}$, can be extracted \cite{Horvatic:2007qs} reasonably 
reliably from the lattice \cite{Alles:2004vi,Alles:1996nm}. 
Thus, it was not modeled in Ref. \cite{Horvatic:2007qs}. 
Hence our assertion that the results of Ref. \cite{Horvatic:2007qs}
unavoidably imply that the straightforward extension of WVR to $T>0$ 
is falsified by experiment \cite{Csorgo:2009pa,Vertesi:2009wf}, 
especially if one recalls that even the sizeable $T$-rescaling 
\cite{Fukushima:2001hr,Schaffner-Bielich:1999uj}
$T_{\rm YM}\to T_{\rm Ch}$ was among the attempts to control 
the $\eta'$ mass enhancement \cite{Horvatic:2007qs}.

Nevertheless, we have also shown that there is a plausible way 
to avoid these problems of the straightforward, naive extension
of WVR to $T>0$, and this is the main result of the present paper. 
Thanks to the existence of another relation, Eq. (\ref{chitilde}), 
connecting the YM quantity $\chi_{\rm Y\!M}$ with QCD quantities 
$\chi$ and $\langle {\bar q}q \rangle_0$, it is possible to define 
a quantity, ${\widetilde \chi}$, which can meaningfully replace 
$\chi_{\rm Y\!M}(T)$ in finite-$T$ WVR. It remains practically 
equal to $\chi_{\rm Y\!M}$ up to some 70\% of $T_{\rm Ch}$, 
but beyond this, it changes following the $T$-dependence of 
$\langle {\bar q}q \rangle_0(T)$. In this way, the successful 
zero-temperature WVR is retained, but the (partial) restoration 
of U$_A$(1) symmetry [understood as the disappearing contribution 
of the gluon anomaly to the $\eta'$ ($\eta_0$) mass] is naturally 
tied to the restoration of the SU$_A$(3) flavor chiral symmetry 
and to its characteristic temperature $T_{\rm Ch}$, instead of 
$T_{\rm YM}$. 

It is very pleasing that this fits in nicely with the recent 
{\it ab initio} theoretical analysis using functional methods 
\cite{Alkofer:2011va}, which finds that the anomalous breaking 
of U$_A$(1) symmetry is related to DChSB (and confinement) in a 
self-consistent manner, so that one cannot have one of these 
phenomena without the other.

Of course, the most important thing is that this version of 
the finite-$T$ WVR, obtained by 
$\chi_{\rm Y\!M}(T)\to {\widetilde \chi}(T)$, is consistent
with experiment \cite{Csorgo:2009pa,Vertesi:2009wf} for all 
reasonable strengths of $T$-dependence [$0\leq \delta \alt 1$
in Eq. (\ref{chi_small_mT})]. Namely, the both tablets in 
Fig.  \ref{M_Pchi1757} show, first, that $\eta'$ mass close to 
$T_{\rm Ch}$ suffers the drop of more than 200 MeV with respect 
to its vacuum value. This satisfies the minimal experimental 
requirement abundantly. Second, 
Fig. \ref{M_Pchi1757} shows an even larger drop of the $\eta_0$ 
mass, to some 400 MeV, close to the ``best" value of the in-medium
$\eta'$ mass (340 MeV, albeit with large errors) obtained by 
Cs\"org\H{o} {\it et al.} \cite{Csorgo:2009pa,Vertesi:2009wf}. 
This should be noted because the $\eta_0$ mass inside the fireball
is possibly even more relevant.  Namely, although it is, strictly 
speaking, not a physical meson, $\eta_0$ is the state with the 
$q\bar q$ content closest to the $q\bar q$ content of the 
physical $\eta'$ {\it in the vacuum}. Thus, among the isoscalar 
$q\bar q$ states inside the fireball, $\eta_0$ has the largest 
projection on, and thus the largest amplitude to evolve,
by fireball dissipation, into an $\eta'$ in the vacuum.

\vskip 3mm

{\bf Acknowledgments}

D. H., D. Kl. and S. B. were supported through the 
project No. 0119-0982930-1016,
and D. Ke. through the project No. 098-0982887-2872
of the Ministry of Science, Education and Sports of Croatia. 
D. H. and D. Kl. acknowledge discussions with R. Alkofer. 
The support by CompStar network is also acknowledged. 

%\vspace{-2mm}

\end{document}